\title{Building benchmarking frameworks for supporting replicability and reproducibility:  spatial and textual analysis as an example}
\author{
        Yingjie Hu \\
                GeoAI Lab, Department of Geography, University at Buffalo, NY 14260, USA       
}
\date{}

\documentclass[11pt]{article}

\usepackage[round]{natbib}
\usepackage{graphicx}
\usepackage[margin=1in]{geometry}
\usepackage{setspace}
\begin{document}
\maketitle




\vspace*{-0.5cm}
Replicability and reproducibility (R\&R) are critical for the long-term prosperity of a scientific discipline. 
In GIScience, researchers have discussed R\&R related to different research topics and  problems, such as local spatial statistics, digital earth, and metadata \citep{fotheringham2009problem,goodchild2012future,anselin2014metadata}. This position paper  proposes to further support R\&R by building benchmarking frameworks in order to facilitate the replication of previous research for effective and efficient comparisons of methods and software tools developed for addressing the same or similar problems. Particularly, this paper will use \textit{geoparsing}, an important research problem in spatial and textual analysis, as an example to explain the values of such benchmarking frameworks.

Today's Big Data era brings large amounts of unstructured texts, such as Web pages, historical archives, news articles, social media posts, incident reports, and business documents, which contain rich geographic information. Geoparsing is a necessary step for extracting structured geographic information from  unstructured texts \citep{doi:10.1080/13658810701626343}. A developed geoparsing system, called a \textit{geoparser}, can take unstructured texts  as the input and output the recognized place names and their corresponding spatial footprints. In recent years, geoparsers are playing an increasingly important role in research related to disaster response, digital humanities, and others.       


Since a number of geoparsers have already been developed by previous studies, a researcher, who would like to propose a new (and better) geoparser, would ideally replicate previous research and compare his or her geoparser with the existing ones in order to demonstrate its superiority. In reality,  conducting such a comparative experiment is often difficult, due to several reasons: \textbf{(1)} Some existing geoparsers do not provide source code. In order to perform a comparison, one has to spend a considerable amount of effort to re-implement a previous method. Even when a researcher does so, the implementation could be criticized as not a \textit{correct implementation} if the comparative results seem to favor the new method by the researcher. \textbf{(2)} For geoparsers which provide source codes, it still takes a lot of time and efforts for one to deploy the code and run it over some datasets, and any incorrect configurations can make the replication unsuccessful. \textbf{(3)} Some studies do not share the data used for training and testing the geoparsers. There exist policy restrictions (e.g., Twitter only allows one to share tweet IDs instead of the full  tweet content) and privacy concerns that prevent one from sharing  data. \textbf{(4)} For studies that do share data, it still takes considerable amount of time for another research group to find this dataset, download it, understand its structure and semantics, and use it for experiments. Due to these reasons, it becomes difficult to replicate previous geoparsing research in order to conduct a comparative  experiment.


Another factor that affects R\&R is the dynamic nature of the Web. With today's fast technological advancements,  algorithms backing online applications, such as search engines and recommendation systems, can change day by day. Consider a researcher (Let's call her researcher A) who published a paper in 2017, in which she compared her geoparser with the state-of-the-art commercial geoparser from a major tech company, and showed that her geoparser had a better performance. Then in 2018, researcher B repeated the experiment and found that the geoparser developed by researcher A, in fact, performed worse than the commercial geoparser from the company. Does this mean the work of researcher A is not replicable? Probably not. The tech company may have internally changed its algorithm in 2018, and therefore the comparative experiment conducted by researcher B is no longer based on the same algorithm  used in the experiment of researcher A.

This position paper proposes   a benchmarking framework for geoparsing, which is an open-source and Web-based system. It addresses the limitations discussed above with two designs. First, it hosts a number of openly available datasets and  existing geoparsers. In order to test the performance of a new geoparser, one can connect the newly developed geoparser to the system, and run it against the other hosted geoparsers on the same datasets. Testing different geoparsers on the same dataset and testing the same geoparser on different datasets are extremely important, since both our previous experiments and other studies show that the performances of different geoparsers can vary dramatically when given different datasets \citep{hu2014improving,gritta2018s}. Researchers can also upload their own datasets to this benchmarking framework for testing. In addition, since the system itself does not publicly share the hosted datasets, it sidesteps the restrictions from some data sharing policies. In short, this design can reduce the time and efforts that researchers have to spend in implementing existing baselines for conducting comparative experiments. Second, the benchmarking framework enables the recording of scientific experiments. As researchers conduct evaluation experiments on this system, details of the experiments are recorded automatically, which can include the date and time, datasets selected, baselines selected, metrics,  experiment results, and so forth. The benchmarking framework will provide researchers  with a unique id which allows them to search the experiment result. One can even provide such an id in papers submitted to journals or conferences, so that reviewers can check the raw results of the experiments quickly. These experiment records can serve as evidence for  R\&R. If we go back to the previous example,     researcher A can provide such an experiment id to prove that she indeed conducted such an experiment and obtained the reported result.

In conclusion, this position paper proposed to build benchmarking frameworks to support  R\&R in geospatial research. While the discussion focused on geoparsing in spatial and textual analysis, the same  idea can be applied to other geospatial problems, such as land use and land cover classification, to facilitate effective and efficient comparisons of methods. Such a framework also records experiment details and allows the search of previous experiment results. The evaluation results from the benchmarking frameworks are not to replace customized evaluations necessary for particular projects, but to serve as supplementary information for understanding  developed methods.

\small
\bibliographystyle{plainnat}
\bibliography{GIR18_bibliography}

\end{document}